\documentclass[prl,aps,twocolumn,superscriptaddress]{revtex4-1}

\usepackage{graphicx}
\usepackage{float}
\usepackage{latexsym,amsmath,amssymb,bm,euscript}
\usepackage{color}
\usepackage{subfigure}
\usepackage{epstopdf}
\usepackage[colorlinks=true,linkcolor=blue,citecolor=blue]{hyperref}

\renewcommand{\paragraph}[1]{{\par\it #1.---}\ignorespaces}
\usepackage{appendix}

\newcommand{\HE}{H_E}
\newcommand{\HP}{H_P}
\newcommand{\dg}{^{\dagger}}

\renewcommand{\vec}[1]{{\boldsymbol{#1}}}
\newcommand{\op}{\hat}
\newcommand{\id}{\mathbb{I}}

\newcommand{\Tr}{\text{Tr}}
\newcommand{\Spec}{\text{Spec}}

\newcommand{\Diag}{\text{Diag}}

\newcommand{\Span}{\text{Span}}
\newcommand{\fenv}{f_{\mathrm{env}}}
\newcommand{\pr}{^{\prime}}

\newcommand{\ff}{f}

\begin{document}
\title{Reconstructing Entanglement Hamiltonian via Entanglement Eigenstates}
\date{\today}
\author{W. Zhu}
\email{weizhu@lanl.gov}
\affiliation{Theoretical Division, T-4, Los Alamos National Laboratory, Los Alamos, NM 87545, USA}

\author{Zhoushen Huang}
\email{zsh@lanl.gov}
\affiliation{Theoretical Division, T-4, Los Alamos National Laboratory, Los Alamos, NM 87545, USA}

\author{Yin-Chen He}
\email{yinchenhe@perimeterinstitute.ca}
\affiliation{Department of Physics, Harvard University, Cambridge MA 02138}
\affiliation{Perimeter Institute for Theoretical Physics, Waterloo, Ontario N2L 2Y5, Canada}

\begin{abstract}
The entanglement Hamiltonian $\HE$, defined through the reduced density matrix of a subsystem $\rho_A=\exp(-\HE)$,
is an important concept in understanding the nature of quantum entanglement
in many-body systems and quantum field theories.
In this work, we explore a numerical scheme which explicitly reconstructs the entanglement Hamiltonian
using one entangled mode (i.e., an eigenstate) of $\rho_A$.
We demonstrate and benchmark this scheme on quantum spin lattice models.
The resulting $\HE$ bears a form similar to a physical Hamiltonian with spatially varying couplings, which allows us to make quantitative comparison with perturbation theory and conformal field theory.
\end{abstract}


\maketitle


\paragraph{Introduction}
Entanglement-based analysis has brought new insights into
the study of condensed matter systems, particularly those with strong
interactions \cite{Amico2008,Laflorencie2016,Eisert2010}, where the understanding
of ground state correlations is of central importance.  Given a pure
state $|\psi\rangle$, the entanglement between two complementary parts
($A$ and $B$) can be extracted from the reduced density matrix of
(say) part $A$,
$\rho_A = \mbox{Tr}_{B} |\psi\rangle \langle \psi|$.
The entanglement entropy is $S = -\Tr(\rho_A \ln \rho_A)$
and has been extensively used to identify quantum criticality
\cite{Vidal2003,calabrese2004entanglement} and intrinsic topological orders \cite{Kitaev2006,
  Levin2006}. Following Li and Haldane \cite{Haldane2008}, more recent
developments have gone beyond the single number $S$, and invoked the
full spectrum $\{p_n\}$ of $\rho_A$, i.e., the entanglement spectrum
(ES), as more fine-grained ``fingerprints'' to distinguish between
various topological orders
\cite{Haldane2008,Lauchli2010,Cincio2013,Zaletel2013}, symmetry
protected phases \cite{Pollmann2010,Pollmann2012}, symmetry broken
phases \cite{Metlitski2011,Alba2013}, quantum criticality
\cite{Chiara2012,Poilblanc2010,Cirac2011,Hsieh2014,Bayat2014,Tommaso2016}, to name a few.

The reduced density matrix can be formally written as
$\rho_A = e^{-\HE}$, and regarded as a thermal density matrix with
``Hamiltonian'' $\HE$ (or entanglement Hamiltonian EH) at inverse
temperature $\beta = 1$. Knowledge of $\HE$ in terms of its operator
content 
could then offer an alternative picture of how subsystem $A$ behaves, by appealing to our intuition of thermodynamics.
Specifically, a concrete form of $\HE$ may provide insight for interesting problem such as bulk-edge correspondence~\cite{Haldane2008,XLQi2012} and physics of thermalization in the non-equilibrium dynamics~\cite{CardyTonni,Cardy2005,XuedaWen}.
From an information extraction point of view, both the entropy and the ES
represent ways of reducing the full information content in $\rho_A$ to
more manageable forms. The reconstruction of the EH, if achievable,
points to a different reduction scheme, whereby the exponentially many
complex-valued matrix elements in $\rho_A$ are compressed into a
handful of coupling constants in $\HE$.
In a limited number of tractable cases, $\HE$ has been explicitly obtained either exactly  \cite{Chung2000,Peschel2003,Fidkowski2010,Klich2017,Peschel2018} or perturbatively  \cite{Lauchli2012}.
To date, however, there is no generic recipe to derive $\HE$ in such a compact form in strongly-correlated systems.

\begin{figure}[t]
  \includegraphics[width=0.45\textwidth]{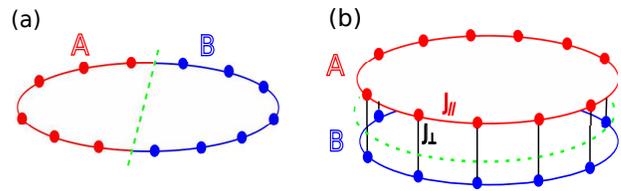}
  \caption{ (a) One-dimensional spin $-1/2$ chain with $2 L$ sites.
  (b) Spin ladder model made of two coupled periodic spin$-1/2$ chain with total $2 L$ sites.
    $J_{\parallel}=\cos\theta$ and $J_{\perp}=\sin\theta$ respectively describes intra-chain and inter-chain couplings.
    The dashed line shows the entanglement bipartition into two subsystem A (red) and B (blue),
    each of which encloses $L$ sites.
  } \label{fig:model}
\end{figure}

In this work, we present a systematic strategy to obtain $\HE$. Instead
of evaluating $-\log \rho_A$ directly, we construct $\HE$ from an eigenstate (practically chosen as the highest weight one) of $\rho_A$.
We consider an ansatz
$\HP = \sum_a w_a L_a$, which is a weighted sum of a prescribed set of physically motivated local operators $\{L_a\}$.
Examples of such operators include spin-spin interactions, fermion hoppings, etc.
The coefficients $\{w_a\}$ will be determined by demanding the
highest weight eigenstate of $\rho_A$ (i.e., the entanglement ground
state) to be an (approximate) eigenstate of $\HP$, which leverages a
method reported in recent works on parent Hamiltonian construction
\cite{XLQi2017,Clark2018,Greiter2018} (see \cite{sm} for an alternative perspective).
Generically, the EH $\HE$ should be a function of $\HP$, $\HE = \ff(\HP)$.
In many physical interesting situations, however, $\HE$ is believed to contain only local terms.
In those cases, one can always choose the local operators $\{L_a\}$ properly such that $\HP$ \emph{is} the EH (with a proper rescaling).
We demonstrate our method using two exemplary spin-$\frac{1}{2}$ models (see Fig.~\ref{fig:model}).
In both models, we obtain numerically exact EHs, which also converge to
analytical forms, if obtainable in the corresponding conformal field
theory (CFT) or perturbatively around exactly solvable points. Toward
the end, we will briefly discuss its implications in Haldane conjecture and
non-equilibrium statistical mechanics.

\paragraph{Method}
Our aim is to obtain the EH,
$\HE = -\log \rho_A$, explicitly in terms of intelligible
operators. This problem is in general analytically untractable due to
the difficulty in evaluating the $\log$. Below, we will instead
(1) Confine ourselves to a restricted operator space $\mathcal{L}$ consisting of linear combinations of a prescribed set of basis operators, $\mathcal{L} = \Span\{L_a\}$, and then
(2) Construct an operator $\HP \in \mathcal{L}$, such that it (approximately) shares one eigenstate with $\rho_A$ (the highest weight state).
In principle, $\HP$ thus constructed may not be $\HE$, instead it could be a certain function of $\HE$.
However, the EH obtained from the groundstate of a local Hamiltonian is itself believed to be local.
Thus as long as one chooses the operators $\{L_a\}$ properly (e.g. by including enough local operators), $\HP$ and $\HE$ should be equivalent up to a proper rescale.
We find this is indeed the case in the two examples to be discussed later.

To obtain $\HP$, we use a recently reported method
\cite{XLQi2017,Clark2018,Greiter2018} which takes as input a state
$|\xi\rangle$ and a set of basis operators $\{L_a\}$, and returns a
set of weights $\{w_a\}$, such that $\HP = \sum_a w_a L_a$ has
$|\xi\rangle$ as an (approximate) eigenstate. Specifically, we take
$|\xi\rangle$ as the entanglement ground state, and compute the
correlation matrix
\begin{gather}
  \label{Gab-def}
  G_{ab} = \langle \xi | L_a L_b | \xi\rangle - \langle \xi | L_a |
  \xi\rangle\langle \xi | L_b | \xi\rangle\ .
\end{gather}
Note that $G$ is positive-semidefinite \cite{sm}. The desired weights are given
by the eigenvector of the matrix $G$ with the lowest eigenvalue
$g_0 \ge 0$,
\begin{gather}
  \{w_a\}: \sum_b G_{ab} w_b = g_0 w_a \ , \ g_0 = \min\{\Spec(G)\} \ge 0 \\
  \HP = \sum_a w_a L_a\ .
\end{gather}
One can easily verify that
$g_0 = \langle \xi | \HP^2 | \xi\rangle - \langle \xi
|\HP|\xi\rangle^2$, \emph{i.e.}, $g_0$ is the ``energy fluctuation'' of
the state $|\xi\rangle$ under ``Hamiltonian'' $\HP$. $|\xi\rangle$
becomes an exact eigenstate of $\HP$ if $g_0 = 0$. For small but
nonvanishing $g_0$, $\HP$ is the best approximate parent ``Hamiltonian''
of $|\xi\rangle$ \footnote{It is the best in the sense that if the
  basis operators $\{L_a\}$ are orthonormal (that is,
  \unexpanded{$\Tr(L_a\dg L_b) = \delta_{ab}\Tr \id$}, where $\id$ is
  the identity), and if \unexpanded{$\sum_a |w_a|^2 = 1$} is also
  normalized, then out of all \emph{normalized} operators
  \unexpanded{$\{H \in \mathcal{L} | \Tr(H\dg H) = \Tr \id\}$}, $\HP$
  has the smallest fluctuation with respect to the given state
  \unexpanded{$|\xi\rangle$}. See \cite{sm} for more details}.

Although the above construction formally only ensures that $\HP$ and
$\HE$ (approximately) share one eigenstate $|\xi\rangle$, we found in
our study that the remainder of the eigenbasis also match well
whenever $g_0$ is small, which we will quantify in the examples later.
Note also that there is no
\emph{a priori} relation between the spectra of $\HP$ and $\HE$ even
when the eigenbases match exactly, this is why we take the more
general form $\HP = \ff(\HE)$.

With $\HP$ fixed, we can determine the best $\ff$, in principle, by
maximizing the density matrix fidelity \cite{Uhlmann76} between the
original $\rho_A$ and its reconstruction $\varrho = e^{-\ff(\HP)}$,
\begin{gather}
  F(\rho_A, \varrho) = \Tr \sqrt{\sqrt{\rho_A}\varrho \sqrt{\rho_A}}\ .
\end{gather}
We write the eigen-decomposition of $\varrho$ as
\begin{gather}
  \varrho(\vec q) = \sum_n q_n |\phi_n\rangle\langle \phi_n|\ ,
\end{gather}
where $\vec q = (q_1, q_2, \cdots)$, $q_n = e^{-\ff(\varepsilon_n)}$,
and $|\phi_n\rangle$ and $\varepsilon_n$ are the $n^{th}$ eigenstate
and eigenvalue of $\HP$, respectively. In SM \cite{sm}, we show that maximizing
$F(\rho_A, \varrho)$ leads to a self-consistent equation of $\vec
q$. Its solution implicitly defines the $\ff$ function through
$\ff(\varepsilon_n) = -\log q_n$.  When the eigenbasis
$\{|\phi_n\rangle\}$ of $\HP$ matches well with the entanglement
states $\{|\xi_n\rangle\}$, the optimal $\vec q$ can be approximated
by (see \cite{sm})
\begin{gather}\label{eq:qn}
  q_n \simeq \langle \phi_n | \rho_A | \phi_n\rangle \ \forall n\ .
\end{gather}
In other words, under this approximation, $\varrho$ describes the
diagonal ensemble of $\rho_A$ in the reconstructed
$\{|\phi_n\rangle\}$ basis.

Before going into examples, we remark that the EH can in principle be calculated by numerically evaluating $-\log \rho_A$ using exact diagonalization.
Such calculations, however, require keeping track of the coefficients of exponentially many operators $|n\rangle\langle n\pr|$ in a manybody complete basis $\{|n\rangle\}$.
Our method is numerically more efficient although it needs extra input regarding the physical properties of the system (reflected in the choice of $\{L_a\}$).
More importantly, our method applies to situations (e.g.~in simulations using matrix product state) where  $\log \rho_A$ is hard to calculate numerically.

\paragraph{One-dimensional chain}
As a first case example, we study the EH of bipartition of a one dimensional spin$-1/2$ chain model (as shown in Fig.~\ref{fig:model}(a)):
\begin{eqnarray*}
  &&\hat H= \sum_{n=1}^{2L} \hat h_{n,n+1}= \sum_{n=1}^{2L} S^x_{n} S^x_{n+1}+S^y_{n} S^y_{n+1} + \Delta S^z_{n} S^z_{n+1}.
\end{eqnarray*}
For $|\Delta|\leq 1$, the ground state can be effectively described by a gapless Luttinger liquid.
Importantly, this phase is an example of quantum critical phase with conformal invariance,
governed by a (1+1) conformal field theory (CFT).
As a benefit of the conformal invariance, the EH
can be directly mapped out \cite{Bisognano1975,Bisognano1976,CasiniHuertaMyers,SRyu2016,CardyTonni}:
\begin{eqnarray} \label{eq:CFT}
  &&\HE^{\mathrm{CFT}}= \sum_{n=1}^{L} \fenv(\tilde n) \hat h_{n,n+1},
\end{eqnarray}
where $\fenv(\tilde{n})=\tilde{n}(1-\tilde{n})$ is the envelope function
and $\tilde n = (n + \frac{1}{2})/L$.

\begin{figure}[b]
  \includegraphics[width=0.49\linewidth]{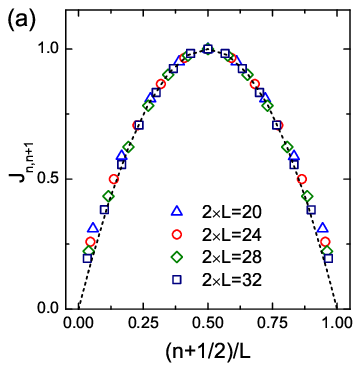}
  \includegraphics[width=0.49\linewidth]{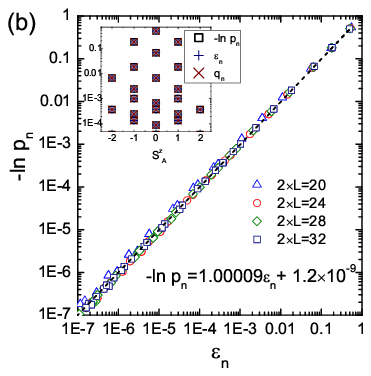}
  \caption{(a) Spatially varying coupling strengths $J_{n,n+1}$ of the parent Hamiltonian of bipartition of a one-dimension spin$-1/2$ chain model.
    The black dashed lines show the CFT predicted envelope function $\fenv(\tilde{n})=\tilde{n}(1-\tilde{n})$ and $\tilde{n}=\frac{n+(n+1)}{2L}$.
    (b) Direct comparison $-\log p_n$ of the ES, $q_n$ by Eq.~\ref{eq:qn} and the eigenvalues $\varepsilon_n$ of parent Hamiltonian $H_P$.
    The black line represents the best linear fit, with the slope $\sim 1.000090$ and intercept $\sim 10^{-9}$.
    Inset: One-to-one comparison of $-\log p_n$ of the ES and eigenvalues $\varepsilon_n$ grouped by quantum number $S^z_A$ in subsystem A.
    Different symbols show the results computed on $2\times L$ systems: blue triangular ($L=10$),
    red circles ($L=12$), green diamonds ($L=14$), navy squares ($L=16$).
  } \label{fig:CFT}
\end{figure}

By implementing the numerical scheme discussed in the method section,
we search for a parent Hamiltonian with the form $H_P= \sum_{n} J_{n,n+1}\hat h_{n,n+1}$ on modest partition sizes.
First of all, we identify one exact zero eigenvalue ($g_0<10^{-13}$) in the spectrum of correlation matrix (Tab.~\ref{tab:CFT}).
The coefficients $J_{n,n+1}$ in $H_P$ can be obtained from the corresponding eigenvector.
Since subsystems A and B both have open boundaries after bipartition (Fig.~\ref{fig:model}(b)),
translation symmetry is broken and $J_{n,n+1}$ is expected to be spatially dependent.
In Fig.~\ref{fig:CFT}(a), we show the spatial dependence of $J_{n,n+1}$ (with proper normalization),
where $J_{n,n+1}$ is non-uniform and takes smaller values near the virtual boundary.
In particular, the dependence of $J_{n,n+1}$ on $n$ (the distance from the boundary) matches the CFT predicted envelope function
$\fenv$ (black dashed line).

The agreement between $J_{n,n+1}$ and $\fenv(\tilde n)$ suggests that
for this model, $H_P$ and $\HE$ are equivalent up to shift
and rescaling, $\HE = \ff(\HP) = a + b\HP$. To verify, we first compare the ES $\{-\log p_n\}$ and
the eigenvalues $\{\varepsilon_n\}$ of $H_P$.  As shown in
Fig.~\ref{fig:CFT}(b), down to order $10^{-7}$, the ES is extremely
well captured by $\{\varepsilon_n\}$ through a simple linear fit,
$-\log p_n = a + b\varepsilon_n$.
Using the fitted $a$ and $b$, we compute the fidelity between the
original and reconstructed RDMs, $F(\rho_A, \varrho)$ where
$\varrho = e^{-(a+b\HP)}$.
As shown in Tab.~\ref{tab:CFT}, $F(\rho_A,\varrho) > 0.9999$ for all
system sizes tested. We thus conclude that $\HP$ and $\HE$ are indeed
equivalent.

\begin{table}[t]
\caption{Lowest eigenvalue $g_0$ of correlation matrix $G$, and density
matrix fidelity $F(\rho_A,\varrho)$ obtained on different system sizes $L$.
Here we set $\Delta=0$ in one-dimension spin$-1/2$ chain model.}
  \begin{tabular}{c|c|c|c|c}
    \hline
    \hline
     $2\times L$ & $20$ & $24$ & $28$ & $32$  \\
    \hline
    $g_0$ &$2.9\times 10^{-14}$& $1.4\times 10^{-13}$& $1.2\times 10^{-13}$ & $9.4\times 10^{-15}$  \\
    $F(\rho_A,\varrho)$ & $0.99999$ & $0.99998$& $0.99996$ & $0.99991$    \\
    \hline
  \end{tabular}\label{tab:CFT}
\end{table}

\paragraph{Spin ladder model}
We turn to study a two-leg spin$-1/2$ ladder Hamiltonian (as shown in Fig.~\ref{fig:model}(b)):
\begin{eqnarray}\label{eq:ham}
  &&\hat H= \hat H_A + \hat H_B +\hat H_{AB} \nonumber\\
  &&\hat H_{\alpha=A(B)}= J_{\parallel} \sum_{\langle ij\rangle} [S^x_{i,\alpha} S^x_{j,\alpha}+S^y_{i,\alpha} S^y_{j,\alpha} + \Delta S^z_{i,\alpha} S^z_{j,\alpha}] \nonumber\\
  &&\hat H_{AB}= J_{\perp} \sum_{i} [S^{x}_{i,A} S^{x}_{i,B} + S^{y}_{i,A} S^{y}_{i,B} + \Delta S^z_{i,A} S^z_{i,B}],
\end{eqnarray}
where $J_{\parallel}=\cos\theta$ describes the nearest-neighbor exchange interaction in each chain,
and $J_{\perp}=\sin\theta$ is ``rung'' exchange coupling between two chains.
Below, we focus on the isotropic case $\Delta=1$ and antiferromagnetic intra-chain coupling $J_{\parallel}>0$ (see \cite{sm} for the anisotropic $\Delta > 1$).
The nature of the ground state
depends on the sign of $J_{\perp}$.
For antiferromagnetic  $J_{\perp}>0$, spin singlets form on the rungs and the ground state
can be viewed as the product of rung singles \cite{Dagotto1993}.
For ferromagnetic $J_{\perp}<0$, the ladder system can be effectively mapped onto a spin$-1$ chain,
thus the ground state is in the ``Haldane'' phase \cite{Haldane1983,Millis1992}.

We now reconstruct the EH $\HE$ on chain $A$ 
using translationally invariant Heisenberg couplings,
\begin{eqnarray} \label{eq:EH2}
  &&\HE= \sum^{N_r}_{n=1} J_n \hat h_n ,\,\,\,\,\,\, \hat h_n =\sum_{i=1}^{L} S_{i}\cdot S_{i+n} ,
\end{eqnarray}
where $\hat h_n$ is the $n$-th neighbor coupling, and
$N_r$ is long-range interaction cut-off.
As before, the coefficients $J_n$ are obtained through diagonalization of correlation matrix $G$.
We identify one \textit{approximate}
zero mode in the correlation spectrum. Tab.~\ref{tab:Hei} shows one typical example of
the corresponding coupling constants in the EH.
First of all, we found that the reconstructed EH is dominated by the nearest-neighbor coupling, $J_1\gg J_{n>1}$.
Further-neighbor couplings decay as inter-spin distance increases, and
we truncated at $N_r = 4^{th}$ neighbor coupling, which already yields very good reconstruction fidelity of $F(\rho_A,\varrho)>0.998$.
The vanishingly small long-ranged interactions reflects locality of the EH.
We thus conclude that the main feature of the EH is captured by a spin$-1/2$ chain
with nearest neighbor antiferromagnetic Heisenberg couplings.
In addition, in Tab.~\ref{tab:Hei}, we observe an unfrustrated ferromagnetic second-neighbor coupling $J_2<0$.
The oscillatory nature of interaction couplings, which can be antiferromagnetic or ferromagnetic depending upon the separation, is reminiscent of the Ruderman-Kittel-Kasuya-Yosida interaction from which indirect interaction couplings in subsystem A can be induced through subsystem B.

\begin{table}[t]
  \caption{Parameters of EH $\HE$ (Eq.~\ref{eq:EH2})
    constructed from the eigenstate of reduced density matrix $\rho_A$.
    Here we set $J_{\perp}/J_{\parallel}=4$ and $\Delta=1$.
  } \label{tab:Hei}
  \begin{tabular}{c|c|c|c|c|c}
    \hline
    \hline
    $2L$&$g_0$&$J_1$&$J_2$&$J_3$&$J_4$\\
    \hline
    $20$ & $9.43\times 10^{-7}$& $0.9979$ & $-0.0642$ & $0.0041$ & $-0.0024$ \\
    $24$ & $6.96\times 10^{-6}$& $0.9979$ & $-0.0646$ & $0.0043$ & $-0.0023$ \\
    $28$ & $5.54\times 10^{-6}$& $0.9979$ & $-0.0647$ & $0.0039$ & $-0.0023$ \\
    $32$ & $4.80\times 10^{-6}$& $0.9979$ & $-0.0637$ & $0.0046$ & $-0.0016$ \\
    \hline
  \end{tabular}
\end{table}

\begin{figure}[t]
  \includegraphics[width=0.5\textwidth]{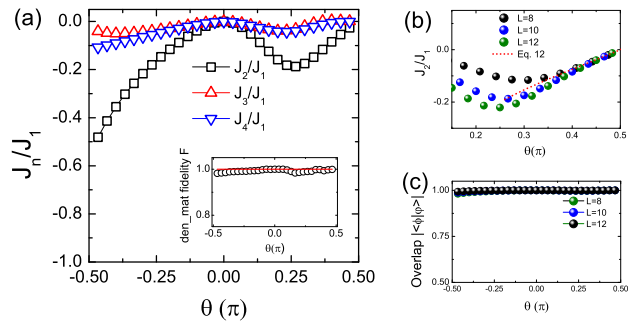}
  \caption{(a) EH parameters $J_n/J_1$ versus $\theta$ (up to fourth nearest neighbor) for a given size $L=12$.
  Inset: Density matrix fidelity $F(\rho_A,\varrho)$ between $\rho_A$ and $\varrho$ versus $\theta$.
  (b) Comparison EH parameters $J_2/J_1$ and the perturbation theory Eq.~\ref{eq:perturbation} (red dashed line).
   Different symbols stand for $L=8$ (green), $L=10$ (blue), $L=12$ (black).
  $\theta\rightarrow \pi/2$ relates to the strong inter-chain coupling limit.
    (c) Wave function fidelity $|\langle \phi^0_E|\varphi^0_A \rangle|$ as a function of $\theta$,
    where $|\varphi^0_A\rangle$ and $|\phi^0_E\rangle$ is the ground state of $\hat H_A$ and $\HE$, respectively.
 } \label{fig:overlap}
\end{figure}

To further understand the obtained EH,
we make a perturbative calculation \cite{sm} in the strong inter-chain coupling limit (${J_{\parallel}\over  J_{\perp}} \ll 1$).
Up to order $\mathcal{O}(\big(\frac{J_{\parallel}}{J_\perp}\big)^2)$, the EH is
         \begin{equation}\label{eq:perturbation}
         H^{\mathrm{per}}_E\approx  J^{\mathrm{per}}_1\sum_i S_i \cdot S_{i+1} - J^{\mathrm{per}}_2 \sum_i S_i \cdot S_{i+2},
         \end{equation}
where $J^{\mathrm{per}}_1=2\frac{J_{\parallel}}{J_{\perp}}$ and $J^{\mathrm{per}}_2=\frac{1}{2}\big(\frac{J_{\parallel}}{J_{\perp}}\big)^2$.
Thus up to $\mathcal{O}(\frac{J_{\parallel}}{J_{\perp}})^2$, subsystem $A$ behaves effectively as a
spin$-1/2$ chain with first- and second-neighbor couplings.
In particular, the second-neighbor coupling is ferromagnetic,  consistent with our results in Tab.~\ref{tab:Hei}.
Fig.~\ref{fig:overlap}(b) shows quantitative agreement between perturbative and numerical results near $J_{\parallel}/J_{\perp} \rightarrow 0$,
where numerics from different system sizes converge to the same perturbation theory values.
This agreement not only provides an analytical
understanding of the oscillatory nature of interaction couplings,
but also validates the accuracy of our numerical results.

One major advantage of our current scheme is
its applicability in the whole parameter regime, which is beyond the
reach of perturbation-based effective theories.
In Fig.~\ref{fig:overlap}(a), we show the EH parameters
as a function of $\theta = \tan^{-1}(J_{\perp}/J_{\parallel})$, up to fourth-neighbor couplings.
At $\theta=0$, the two chains are effectively decoupled,
thus it is reasonable to obtain  $J_{n>1}$ tending to zero.
Away from this decoupling point, generally long-ranged interaction terms appear in $\HE$.
We note that the obtained couplings $J_n/J_1$ show non-monotonic dependence on $\theta$.

With the reconstructed EH in hand, a natural question is if it belongs in the same class with its physical counterpart $\op H_A$.
Since the ground state of $\HE$ can be smoothly and adiabatically connected to that of $\op H_A$ without gap closing (Fig.~\ref{fig:overlap}(c)),
we conclude that $\HE$ and $\op H_A$ are indeed in the same 
class
\footnote{Here, we use the definition that two Hamiltonians are in the same class,
if they can be smoothly connected  through a family of Hamiltonians}.
Interestingly, even though the whole system experiences a quantum phase transition at $\theta=0$,
the EH still faithfully represents the physical Hamiltonian $\op H_A$.

\paragraph{Summary and Discussion}
We have presented a numerical scheme to reconstruct
the entanglement Hamiltonian $\HE$ based on entangled modes
of reduced density matrix, with the help of the recently reported eigenstate-to-Hamiltonian mapping \cite{XLQi2017,Clark2018,Greiter2018}.
As a proof of principle, we applied this method to two quantum spin lattice models.
We found that the reconstructed $\HE$ accurately recovers the expected results and
faithfully captures all features of the reduced density matrices, which are
evidenced by direct comparison to analytical theories,
the agreement between the original and reconstructed full entanglement spectra, and the close-to-$1$ density matrix fidelity.

This scalable recipe for constructing the entanglement Hamiltonian
opens up a number of directions worthy of further exploration.
We explicitly showed in our examples that $\HE$ 
bears a similar form as the physical Hamiltonian,
which unambiguously supports the conjecture that there exists a deep correspondence between
the entanglement Hamiltonian and the physical Hamiltonian with a virtual boundary \cite{Haldane2008,XLQi2012}.
Similar numerical calculations may be used to investigate the time evolution of entanglement Hamiltonian after a quantum quench~\cite{CardyTonni,Cardy2005,XuedaWen}, which may provide intuitive pictures and additional insights
regarding the nature of entanglement propagation and subsystem thermalization.
This work also paves the way for future studies of entanglement Hamiltonian in higher dimensions using matrix product state and similar variational ansatz, for which the correlation matrix (Eq.~\ref{Gab-def}) remains accessible at intermediate system sizes.

\paragraph{Note added} At the final stage of preparing this manuscript,
we became aware of a different scheme to map out entanglement Hamiltonian \cite{Assaad2018}.

\begin{acknowledgments}
\paragraph{Acknowledgments}
W.Z. thanks Y. Zhang and D. N. Sheng for fruitful discussion.
We also thank Xueda Wen for stimulating discussion. Z.S.H. thanks J.-X. Zhu for related discussions.
This work was supported by U.S. DOE at Los Alamos National Laboratory (W.Z., Z.S.H.).
Y.C.H. was supported by the Gordon and Betty Moore Foundation under the EPiQS initiative, GBMF4306, at Harvard University.
This research was also supported in part by Perimeter Institute for Theoretical Physics (Y.C.H.). Research at Perimeter Institute is supported by the Government of Canada through the Department of Innovation, Science and Economic Development Canada and by the Province of Ontario through the Ministry of Research, Innovation and Science.
\end{acknowledgments}

\bibliographystyle{apsrev}
\bibliography{entangled_hamiltonian}{}

\clearpage
\begin{widetext}
  \widetext

  \appendix
  \begin{appendices}

  \tableofcontents

\section{I. Constructing a parent operator from an eigenstate, a linear dependency perspective}
In Ref.~\onlinecite{XLQi2017} (see also
Refs.~\onlinecite{Clark2018,Greiter2018}), Qi and Ranard showed that
given a manybody wavefunction $|v\rangle$, a (more or less)
unique parent Hamiltonian can be constructed in the form
\begin{gather}
  \label{h-lin}
  H = \sum_i w_i L_i\ ,
\end{gather}
where $\{L_i\}$ is a set of Hermitian operators, if and only if
the following ``correlation matrix'' $M^{(v)}_{ij}$ has a unique zero
eigenvalue (with eigenvector $(w_1, w_2, \cdots)$),
\begin{gather}
  M^{(v)}_{ij} \equiv \frac{1}{2} \langle v | \{L_i\ , \ L_j\} | v\rangle - \langle v | L_i | v\rangle \langle v | L_j | v\rangle\ , \\
  \sum_j M_{ij}^{(v)} w_j \stackrel{!}{=} 0 \ .
\end{gather}
Restricting $\{L_i\}$ to spatially local operators, the above
observation then provides a guiding principle for constructing a local
parent Hamiltonian for an arbitrary state $|v\rangle$. Note that
$|v\rangle$ is not necessarily the ground state of thus constructed $H$.

We now provide an alternative perspective for the above and other
related results, in terms of a linear dependence analysis.  A
\emph{sufficient and necessary} condition for a normalized state
$|v\rangle$ to be an eigenstate of $H$ is that
\begin{gather}
  \label{eig-cond}
  (\id - P_v) H | v\rangle = 0 \quad ,\quad P_v \equiv
  |v\rangle\langle v|\ ,
\end{gather}
where $\id$ is identity, and $P_v$ projects onto $|v\rangle$. Consider
now a Hamiltonian of the form Eq.~\ref{h-lin}. Then
\begin{gather}
  (\id - P_v) H |v\rangle = \sum_i w_i |u_i\rangle\quad , \quad
  |u_i\rangle \equiv (\id - P_v) L_i |v\rangle\ .
\end{gather}
The unnormalized $\{|u_i\rangle\}$ states are generated by first
``exciting'' $|v\rangle$ by $L_i$, and then projecting out the part
parallel to $|v\rangle$. Eq.~\ref{eig-cond} is equivalent to demanding
that the $\{|u_i\rangle\}$ states are \emph{linearly dependent},
\begin{gather}
  \label{u-lin-dep}
  \sum_i w_i |u_i\rangle = 0\ .
\end{gather}

Linear dependence of a set of vectors can be checked via a principal
component analysis, which is mathematically equivalent to a singular
value decomposition (SVD). To proceed, we construct a $D \times M$
matrix $A$ by arranging $|u_i\rangle$ as its $i^{th}$ column,
\begin{gather}
  \label{A-def}
  A_{D\times M} \equiv ( |u_1\rangle \ , \ |u_2\rangle\ , \ \cdots)\ .
\end{gather}
Here $D$ is the full Hilbert space dimension, and $M$ is the rank of
the operator set $\{L_a\}$, $a = 1,2,\cdots, M$. The linear dependence
condition Eq.~\ref{u-lin-dep} is formally equivalent to demanding that
$A$ has at least one zero \emph{singular} value (a more detailed
discussion of the related SVD will be provided below). Equivalently,
the overlap matrix $G_{ij} = (A\dg A)_{ij} = \langle u_i | u_j\rangle$
should have at least one zero \emph{eigenvalue}, with the coefficients
$\{w_i\}$ given by the corresponding eigenvector,
\begin{gather}
  \label{g-def}
  G_{ij} = \langle u_i | u_j\rangle = \langle v | L_i L_j|v\rangle -
  \langle v | L_i | v\rangle \langle v | L_j | v\rangle\ , \\
  \sum_jG_{ij} w_j \stackrel{!}{=} 0\ .
\end{gather}
Note that Qi and Ranard's correlation matrix $M$ is the real part of
the hermitian $G$ matrix. Replacing $G$ with $M$ is equivalent to
enforcing real-valuedness of the resulting coefficients $\{w_i\}$, as
required by the Hermiticity of $H = \sum w_iL_i$. A non-Hermitian
parent operator $H$ can be viewed as an annihilator of the state
$|v\rangle$, as discussed in Ref.~\onlinecite{Greiter2018}

\subsection{1. Principal component analysis of the states
  $\{L_i|v\rangle\}$}
In practice, the choice of the basis operators $\{L_i\}$ is often
based on physical intuition, so for efficiency reasons one may start
with a relatively small set of $\{L_i\}$, and gradually add in more
operators (e.g., in increasing order of spatial span or other physical
preferences), until the lowest singular value of $A$ (or eigenvalue of
$G = A\dg A$) converges toward zero. A natural question therefore
concerns the meaning of the SVD of $A$, which we now address. The SVD reads
\begin{gather}
  \label{A-svd}
  A_{D\times M} = \textsf L \Lambda \textsf R\dg = \sum_{i = 1}^M \lambda_i |l_i\rangle\langle r_i|\quad , \quad \Lambda = \Diag(\lambda_1, \lambda_2, \cdots, \lambda_M)\ , \\
  \textsf L_{D\times M} = (|l_1\rangle, |l_2\rangle, \cdots, l_M\rangle)\quad
  , \quad \textsf R_{M\times M} = (|r_1\rangle, |r_2\rangle, \cdots,
  |r_M\rangle\ .
\end{gather}
The columns of $\textsf L$ and $\textsf R$ are the left and right
singular vectors, respectively, and are denoted as $|l_i\rangle$ and
$|r_i\rangle$. Note that the right singular vectors (which are the
eigenvectors of $G$) are $M$-dimensional. Vectors in the right
singular space $\Span\{|r_i\rangle\}$ represent operators in the
operator space $\Span\{L_i\}$: Writing the $i^{th}$ right singular
vector as
\begin{gather}
  |r_i\rangle = (r_i^{(1)}, r_i^{(2)}, \cdots, r_i^{(M)})^t\ ,
\end{gather}
then the corresponding ``Hamiltonian'' is
\begin{gather}
  \label{Hi-def}
  H^{(i)} \equiv \sum_j L_j r_i^{(j)} = ( L_1, L_2, \cdots, L_M)
  |r_i\rangle\ ,
\end{gather}
similar in spirit to writing polarized spin operators as
$\sigma_{\vec b} = \vec b \cdot \vec \sigma$. One can then verify that
\begin{gather}
  \label{non-eig-eq}
  (\id - P_v) H^{(i)} |v\rangle = A|r_i\rangle = \lambda_i |l_i\rangle\ .
\end{gather}
The first equality follows from Eq.~\ref{A-def}, and the second one
follows from Eq.~\ref{A-svd}. Note that
$\langle v | l_i\rangle = 0 \forall i$, which can be checked by left
multiplying $\langle v |$ to the above equation.
In words, this equation means that the action of $H^{(i)}$ on
$|v\rangle$ generates a deviation, perpendicular to $|v\rangle$, as
given by the corresponding left singular vector $|l_i\rangle$, with
weight $\lambda_i$ (the singular value). In particular, if
$\lambda_i = 0$, then one recovers Eq.~\ref{eig-cond}, and $|v\rangle$
becomes an eigenstate of $H^{(i)}$. Thinking of
$H^{(i)} = i \partial_t$ as a time evolution generator, then the LHS
is the \emph{covariant} time derivative $iD_t$.  The $i^{th}$ left
singular vector $|l_i\rangle$ is thus the normalized tangent vector
generated by $H^{(i)}$, and the corresponding singular value is
related to the Fubini-Study metric in the time direction,
$\lambda_i^2 = \langle v | D_t^2 | v\rangle$, which is also the energy
fluctuation,
\begin{gather}
  \lambda_i^2 = \langle v | H^{(i)} (\id - P_v) H^{(i)} | v\rangle =
  \langle {H^{(i)}}^2 \rangle_v - \langle H^{(i)}\rangle_v^2\ .
\end{gather}
\subsection{2. In what sense is the reconstructed parent operator optimal?  }
The right singular vectors satisfy orthonormality
$\langle r_i |r_j\rangle = \delta_{ij}$. What does it entail for their
operator counterparts $H^{(i)}$ (Eq.~\ref{Hi-def})?  In order to carry
this over to the operator space, one should additionally require the
operators $\{L_i\}$ to satisfy certain operator orthonormality, which,
up until now, we have not enforced. Following Qi and Ranard
\cite{XLQi2017}, we use the Hilbert-Schmidt inner product for
operators,
\begin{gather}
  \langle A, B\rangle \equiv \frac{1}{\Tr \id} \Tr(A\dg B)\ ,
\end{gather}
where $\Tr \id = D$ is the full Hilbert space dimension. An
orthonormal \emph{operator} basis $\{L_i\}$ satisfies
\begin{gather}
  \langle L_i, L_j\rangle \stackrel{!}{=} \delta_{ij} .
\end{gather}
Then ``Hamiltonians'' corresponding to different right singular
vectors also satisfy orthonormality,
\begin{gather}
  \langle H^{(i)}, H^{(j)}\rangle = \sum_{i\pr, j\pr }
  r_i^{(i\pr)}r_j^{(j\pr)} \langle L^{(i\pr)}, L^{(j\pr)}\rangle =
  \langle r_i | r_j\rangle = \delta_{ij}\ .
\end{gather}
In other words, these ``eigen-Hamiltonians'' $\{H^{(i)}\}$ form an
orthonormal basis for the operator space spanned by $\{L_i\}$. A
normalized \emph{traceless} ``Hamiltonian'' $H$ simply means its
spectrum has unit variance, $\Tr(H^2)/\Tr(\id) \stackrel{!}{=}1$.

Using orthonormal $\{L_i\}$, then in situations where an exact zero
eigenvalue does not exist for $G$ (Eq.~\ref{g-def}), the parent
operator $H^{(i_{min})}$ corresponding to the lowest eigenvalue of $G$
is an ``optimal'' approximate parent Hamiltonian, in the sense that
out of all \emph{normalized} operators in the space of $\Span\{L_i\}$,
$H^{(i_{min})}$ generates the lowest energy fluctuation on
$|v\rangle$, or equivalently the least deviation of $H|v\rangle$ from
$|v\rangle$.

\section{II. Quantifying the quality of the reconstructed basis using IPR}
The method described in the text is based on the ansatz that the RDM
$\rho_A$ can be written as a scalar function $y$ of a local operator
$\HP$, and $\HP$ itself is to be (approximately) constructed, from an
exact eigenstate $|\xi\rangle$ of $\rho_A$, in the space of
$\mathcal{L} \equiv \Span\{L_i\}$,
\begin{gather}
  \label{rho-y-HP}
  \rho_A \stackrel{?}{=}y(\HP)\quad , \quad \HP \in \mathcal{L}\ .
\end{gather}
The construction scheme for $\HP$, however, only guarantees that $\HP$
and $\rho_A$ (approximately) share one eigenstate $|\xi\rangle$, with
no constraint on the remainder of the eigenbasis. Therefore, to claim
that one has successfully reconstructed $\rho_A$ in terms of
$\{L_i\}$, one needs to verify that the entire eigenbasis of
$\HP$ approximately matches that of $\rho_A$.

A simple way to quantify the quality of one set of basis states
$\{|\phi_n\rangle\}$ in terms of their similarity to a reference basis
$\Psi \equiv \{|\psi_n\rangle\}$, is to use the inverse participation
ratio, \newcommand{\IPR}{\text{IPR}}
\begin{gather}
  \IPR(\phi_n | \Psi) = \frac{1}{\sum_{m=1}^N |\langle \phi_n |
    \psi_m\rangle|^4} \in [1,N]\ .
\end{gather}
The IPR measures effectively how many basis states in $\Psi$ one needs
to span a particular $|\phi_n\rangle$. It is $1$ if
$\langle \phi_n | \psi_m\rangle = \delta_{m,n}$, and saturates to $N$
if
$|\langle \phi_n | \psi_m\rangle| = \frac{1}{\sqrt{N}} \ \forall
m$. In the context of RDM reconstruction, one would compute the IPR
for each of the eigenstates of $\HP$ in the exact eigenbasis of
$\rho_A$; if all of them are close to $1$, then $\HP$ and $\rho_A$
approximately share the same set of basis states.

\subsection{1. Generalized IPR in the presence of degeneracy}
When $\HP$ has degeneracy, there is a $U(M)$ indeterminacy in an
$M$-fold degenerate subspace $\mathcal{M}$. Then taking a single
numerically obtained eigenstate out of this $M$-dimensional subspace
may yield a ``broadened'' IPR (i.e., one $> 1$), even if upon a $U(M)$
transformation, each of the $M$ (transformed) states could have a
perfectly sharp IPR (i.e., $=1$). To fix this, we generalize the
notion of IPR to a degenerate subspace. Denote the projection operator
of this subspace and a corresponding density operator as
\begin{gather}
  P_{\mathcal{M}} = \sum_{n = n_1}^{n_M}|\phi_n\rangle\langle \phi_n | \quad , \quad \rho_{\mathcal{M}} = P_{\mathcal{M}} / \Tr P_{\mathcal{M}}\ .
\end{gather}
The generalized IPR is defined as the exponentiated $2^{nd}$ Renyi
entropy, $e^{S_2}$, of the diagonal ensemble in the $\Psi$ basis,
\begin{gather}
  \label{ipr-degen}
  \IPR(P_{\mathcal{M}} | \Psi) = \frac{1}{ \sum_{m=1}^N \langle \psi_m | \rho_{\mathcal{M}} | \psi_m\rangle^2 }\ .
\end{gather}
One can verify that the generalized IPR reduces to the standard one
when there is no degeneracy ($M \rightarrow 1$).  Note that if
$P_{\mathcal{M}}$ exactly matches an equal-dimensional subspace in the
$\Psi$ basis,
$P_{\mathcal{M}} = \sum_{n = n_1}^{n_M}|\psi_n\rangle\langle \psi_n|$,
then
$\langle \psi_m | \rho_{\mathcal{M}} | \psi_m\rangle = \frac{1}{M}$,
hence $\IPR(P_{\mathcal{M}} | \Psi) = M$. In other words, in the
perfect match case, the generalized IPR is given by the dimension of
the degenerate subspace $\mathcal{M}$. On the other hand, if each of
the degenerate $|\phi_{n_i}\rangle$ still satisfies
$|\langle \phi_{n_i} | \psi_m\rangle| = \frac{1}{\sqrt{N}} \forall m$,
then
$\langle \psi_m | \rho_{\mathcal{M}} | \psi_m\rangle = \frac{1}{N}$,
hence $\IPR = N$. The generalized IPR thus reflects the notion of
effective number of $|\psi\rangle$ states needed to span the subspace
$P_{\mathcal{M}}$.

\subsection{2. Direct comparison the eigenstates of entanglement Hamiltonian  and those of parent Hamiltonian using IPR}
In this section, we explicitly show the comparison of eigenstates of $H_P$ with those of $\HE$ using IPR.
We take the 1D spin chain as example again. Since the reconstructed $\HE$ has degeneracy, we use the generalized IPR introduced in Eq.~\ref{ipr-degen} when appropriate.
In Fig.~\ref{fig:IPR}, we show the IPR of eigenstates of $H_P$ as labeled by their (renormalied) weight $\varepsilon_n$.
It is found $IPR_n \approx 1$ for all of eigenstates with weight $\varepsilon_n>10^{-6}$,
showing that each eigenstate of $\HE$ is identical to the eigenstate of $\HP$.
Please note that, for worst case, if the eigenstates of $H_P$ and that of $H_E$ are totally independent,
it should be expected maximum value of $IPR\sim N\sim 2^{L/2}$ ($L$ total system size) which is exponential growing with $L$.
In Fig.~\ref{fig:IPR}, $IPR_n$ are all close to $1$ show that the eigenstates of $H_P$
has well captured the eigenstates of $H_E$.

In the main text, we have demonstrated that the eigenvalue of  parent Hamiltonian $H_P$
has one-to-one correspondence with the entanglement spectra of reduced density matrix.
Here, we further show that each eigenstate of $\HP$ can be captured by the eigenstate of $\HE$.
Taking into account that density matrix fidelity $F(\rho_A, \varrho) = \Tr \sqrt{\sqrt{\rho_A}\varrho \sqrt{\rho_A}}$ itself
reveals the weighted averaged wavefunction overlap between the eigenstates of $H_P$ and that of $H_E$,
we now can understand very large value of density matrix fidelity as shown in the main text.
In conclusion, the very large
density matrix fidelity unambiguously sets up the equivalence between $H_P$ and entanglement Hamiltonian $H_E$.

\begin{figure}
\includegraphics[width=0.5\textwidth]{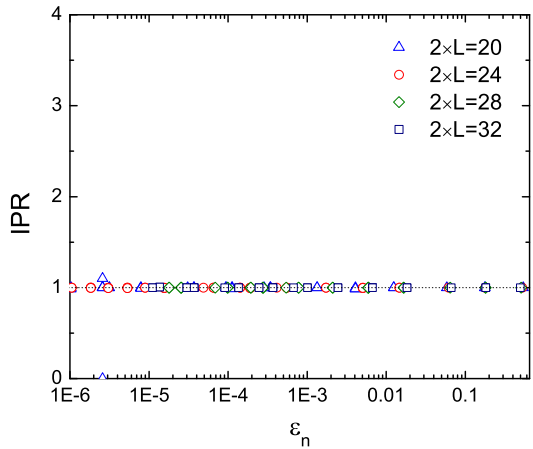}\\
\caption{Inverse participation ratio of eigenstates of parent Hamiltonian $H_P$.
} \label{fig:IPR}
\end{figure}

\section{III. Optimizing RDM reconstruction fidelity}
Under the ansatz Eq.~\ref{rho-y-HP}, if the eigenbasis of the
constructed $\HP$ matches exactly with that of the target RDM
$\rho_A$, then the scalar function $y$ is implicitly determined
through the map between their spectra, $y(\varepsilon_n) = p_n$, where
$\varepsilon_n$ and $p_n$ are the eigenvalue of $\HP$ and $\rho_A$,
respectively, associated with their common eigenvector
$|\psi_n\rangle$. When the basis reconstruction is only approximate,
the best $y$ function can be determined in principle by maximizing the
fidelity between the original and reconstructed RDMs.
For clarity, in this section we will drop the subscript $_A$ and
denote the target RDM as $\rho$. Its eigen decomposition is
\begin{gather}
  \rho = \sum_n p_n |\psi_n\rangle\langle \psi_n|\ .
\end{gather}
The reconstructed density matrix is $\varrho = y(\HP)$ with as of yet
unknown $y$, where
$\HP = \sum_n \varepsilon_n |\phi_n\rangle\langle \phi_n|$ is the
reconstructed parent operator. The eigen decomposition of $\varrho$ is
therefore
\begin{gather}
  \varrho = \sum_n q_n |\phi_n\rangle\langle \phi_n |\ ,
\end{gather}
where $q_n = y(\varepsilon_n)$. The fidelity between the original and
the reconstructed RDMs is defined as
\begin{gather}
  \label{fhat-def}
  F(\rho,\varrho) = \Tr \op F(\rho, \varrho)\quad , \quad \op F(\rho, \varrho) \equiv \sqrt{\sqrt{\rho}\, \varrho \sqrt{\rho}}\ .
\end{gather}
It will be useful to note that the operator $\op F(\rho,\varrho)$ arises from the following \emph{polar decomposition},
\begin{gather}
  \label{polar-decomp}
  \sqrt{\rho_1} \sqrt{\rho_2} = \op F(\rho_1, \rho_2) U(\rho_1,
  \rho_2)\quad , \quad U(\rho_1, \rho_2)\dg = U(\rho_2, \rho_1)\ ,
\end{gather}
where the unitary $U(\rho_1, \rho_2)$, although not of our concern in
the present context, is related to Uhlmann's parallel transport of
density matrices \cite{Uhlmann76}, and the second equation follows
from the hermiticity of $\op F(\rho_1, \rho_2)$. The maximization of
$F$ can be viewed as a variational problem in the space of normalized
distributions $\{q_n\}$, and once the optimal weights are obtained,
$y$ can be determined (or defined) through $y(\varepsilon_n) = q_n$.

The stationary condition for extremal $F$ over the variational space
of $\{q_n\}$ is
\begin{gather}
  \label{max-F-cond}
  \frac{\partial}{\partial q_n} \Bigl[F - \lambda (\sum_m q_m-1)\Bigr] = 0\ ,
\end{gather}
where $\lambda$ is the Lagrangian multiplier for the normalization
$\sum_m q_m = 1$. Using
$\partial \Tr \sqrt{A} = \frac{1}{2} \Tr( \sqrt{A}^{-1} \partial A)$
for any invertible operator $A$, Eq.~\ref{max-F-cond} becomes
\begin{gather}
  \label{stat-cond}
  \frac{\partial F}{\partial q_n} = \frac{1}{2} \langle \phi_n | \op Q |
  \phi_n\rangle = \lambda \quad  \forall n\ ,
\end{gather}
 where
$\op Q = \sqrt{\rho} \op F(\rho,\varrho)^{-1} \sqrt{\rho}$. Using
Eq.~\ref{polar-decomp}, one can show that
$\op F(\rho,\varrho)^{-1} = U(\rho,\varrho) \sqrt{\varrho}^{-1}
\sqrt{\rho}^{-1}$, thus
\begin{gather}
  \op Q = \sqrt{\varrho}^{-1} \op F(\varrho,\rho) \sqrt{\varrho}^{-1}\ ,
\end{gather}
and Eq.~\ref{stat-cond} becomes
$\langle \phi_n | \op F(\varrho, \rho) | \phi_n\rangle =
2\lambda q_n\, \forall n$.  Note that $\sum_n q_n = 1$, thus
$2\lambda = \sum_n LHS = \Tr \op F(\varrho,\rho) = F(\varrho,\rho)$,
and we finally arrive at a self consistent equation for the weights
$\{q_n\}$,
\begin{gather}
  \label{stat-cond-2}
  \frac{\langle \phi_n | \op F(\varrho, \rho) |
    \phi_n\rangle}{\sum_n \langle \phi_n | \op F(\varrho, \rho) |
    \phi_n\rangle} = q_n \quad \forall n\ ,
\end{gather}
note that the LHS depends on $\{q_n\}$ only through $\varrho$.

\subsection{1. Approximate optimal solution in the high-fidelity limit}
When the two bases $\{|\psi\rangle\}$ and $\{|\phi\rangle\}$ have a
good match, the fidelity operator $\op F$ (Eq.~\ref{fhat-def}) is
dominated by its diagonal line (say, in the $\{|\phi\rangle\}$
basis). In this case one may adopt the approximation that
\begin{gather}
  \langle \phi_n | \sqrt{\op F^2(\varrho,\rho)} | \phi_n\rangle \simeq \sqrt{\langle \phi_n | \op F^2(\varrho,\rho)| \phi_n\rangle} = \sqrt{q_n} \sqrt{\langle \phi_n | \rho | \phi_n\rangle}\ .
\end{gather}
Substituting this into Eq.~\ref{stat-cond-2}, one then obtains
\begin{gather}
  q_n \simeq \langle \phi_n | \rho | \phi_n\rangle\ ,
\end{gather}
that is, the optimal $q_n$ is the weight of the reconstructed
eigenstate $|\phi_n\rangle$ in the original (i.e. target) mixed state
$\rho$.

\section{IV. Entanglement Hamiltonian in Strong Inter-chain Coupling Limit}
\label{app:perturbation}
We will derive the entanglement Hamiltonian $\HE$ in the strong inter-chain coupling limit using perturbation theory.
    The starting point is the physical Hamiltonian:
    \begin{eqnarray}\label{eq:ham1}
      &&\hat H= \hat H_A + \hat H_B +\hat H_{AB} \nonumber\\
      &&\hat H_{\alpha=A(B)}= J_{\parallel} \sum_{\langle ij\rangle} [S^x_{i,\alpha} S^x_{j,\alpha}+S^y_{i,\alpha} S^y_{j,\alpha} + \Delta S^z_{i,\alpha} S^z_{j,\alpha}] \nonumber\\
      &&\hat H_{AB}= J_{\perp} \sum_{i} [S^{x}_{i,A} S^{x}_{i,B} + S^{y}_{i,A} S^{y}_{i,B} + \Delta S^z_{i,A} S^z_{i,B}].
    \end{eqnarray}
    In the limit of $J_{\perp}\gg J_{\parallel}$, we treat $\hat H_{A(B)}$ as the perturbation to $\hat H_{AB}$.
    Thus the ground state of $\hat H_{AB}$ can be viewed as a product state of spin singlets:
    \begin{equation}\label{}
      |0\rangle =\prod_i |s_i \rangle,
    \end{equation}
    where $|s_i \rangle$ is the spin singlet living on inter-chain bond:
    \begin{equation}\label{}
      |s_i \rangle = \frac{1}{\sqrt{2}} (|\uparrow_{i,A}\rangle|\downarrow_{i,B}\rangle -|\downarrow_{i,A}\rangle|\uparrow_{i,B}\rangle  ),\,\,\,\,\, E_s=(-\frac{1}{2}-\frac{\Delta}{4}) J_{\perp}
    \end{equation}
    On each inter-chain bond, spin excitation state is described by spin triplet excitations:
    \begin{eqnarray}\label{}
      |t^+_i \rangle &=&  |\uparrow_{i,A}\rangle|\uparrow_{i,B}\rangle ,\,\,\,\,\,\,\,\,\,\,\,\, E_{t^+}= \frac{\Delta}{4}J_{\perp}   \nonumber\\
      |t^0_i \rangle &=& \frac{1}{\sqrt{2}} (|\uparrow_{i,A}\rangle|\downarrow_{i,B}\rangle +|\downarrow_{i,A}\rangle|\uparrow_{i,B}\rangle ),\,\,\,\,\, E_{t^0}=(\frac{1}{2}-\frac{\Delta}{4})J_{\perp} \nonumber\\
      |t^-_i \rangle &=&  |\downarrow_{i,A}\rangle|\downarrow_{i,B}\rangle  ,\,\,\,\,\,\,\,\,\,\,\,\,\, E_{t^-}=\frac{\Delta}{4}J_{\perp}.
    \end{eqnarray}

    At first-order perturbation theory, the first-order correction is
    \begin{eqnarray*}
      |1\rangle &=& \sum_{i} |t^+_i t^-_{i+1}\rangle \frac{\langle t^+_i t^-_{i+1}|\hat H_A+\hat H_B| 0\rangle}{E_{+}+E_{-}-2E_s} +
                    |t^-_i t^+_{i+1}\rangle \frac{\langle t^-_i t^+_{i+1}|\hat H_A+\hat H_B| 0\rangle}{E_{-}+E_{+}-2E_s}+
                    |t^0_i t^0_{i+1}\rangle \frac{\langle t^0_i t^0_{i+1}|\hat H_A+\hat H_B| 0\rangle}{2E_{0}-2E_s}  \nonumber\\
                &=& \frac{J_{\parallel}}{4J_{\perp}} \sum_{i}[
                    \frac{2}{1+\Delta} | t^+_it^-_{i+1}\rangle  + \frac{2}{1+\Delta} | t^-_it^+_{i+1}\rangle - \Delta |t^0_i t^0_{i+1}\rangle  ]
    \end{eqnarray*}
    , where we use the notation:
    \begin{eqnarray*}
      && |t^+_i t^{-}_{i+1} \rangle= |s_{1}\rangle\otimes... |s_{i-1}\rangle | t^+_i\rangle |t^-_{i+1}\rangle \otimes|s_{i+2}\rangle...|s_{L}\rangle\\
      && |t^-_i t^{+}_{i+1} \rangle= |s_{1}\rangle\otimes... |s_{i-1}\rangle | t^-_i\rangle |t^+_{i+1}\rangle \otimes|s_{i+2}\rangle...|s_{L}\rangle\\
      && |t^0_i t^{0}_{i+1} \rangle= |s_{1}\rangle\otimes... |s_{i-1}\rangle | t^0_i\rangle |t^0_{i+1}\rangle \otimes|s_{i+2}\rangle...|s_{L}\rangle
    \end{eqnarray*}
    and Hamiltonian elements can be calculated by using:
    \begin{eqnarray*}
      &&  S^+_{i,A} S^-_{i+1,A} | s_i\rangle | s_{i+1}\rangle = -\frac{1}{2}| t^+_i\rangle |t^-_{i+1}\rangle \\
      &&  S^-_{i,A} S^+_{i+1,A} | s_i\rangle | s_{i+1}\rangle = -\frac{1}{2}| t^-_i\rangle |t^+_{i+1}\rangle \\
      &&  S^z_{i,A} S^z_{i+1,A} | s_i\rangle | s_{i+1}\rangle =  \frac{1}{4}| t^0_i\rangle |t^0_{i+1}\rangle
    \end{eqnarray*}

    The reduced density matrix can be obtained by (within first-order perturbation approximation):
    \begin{equation}\label{}
      \rho_A=Tr_B[|\psi\rangle \langle \psi|]= Tr_B[(|0\rangle+|1\rangle) (\langle 0|+\langle 1|)]
    \end{equation}

    First we get
    \begin{eqnarray*}
      Tr_B |0\rangle\langle 0| &=& \prod_i Tr_B |s_i \rangle \langle s_i| \\
                               &=& \prod_i [\langle \uparrow^B_i|s_i \rangle \langle s_i| \uparrow^B_i\rangle + \langle \downarrow^B_i|s_i \rangle \langle s_i| \downarrow^B_i\rangle] \\
                               &=& \prod_i \frac{1}{2}[ |\uparrow^A_i \rangle\langle \uparrow^A_i| + |\downarrow^A_i \rangle\langle \downarrow^A_i|] = \frac{1}{2^L}
    \end{eqnarray*}

    Second, we calculate
    \begin{eqnarray*}
      Tr_B [|1\rangle\langle 0| +|1\rangle\langle 0|] &=&
                                                          \frac{J_{\parallel}}{4J_{\perp}}[ -\Delta Tr_B |t^0_it^0_{i+1}\rangle \langle 0| + \frac{2}{1+\Delta}Tr_B |t^+_it^-_{i+1}\rangle \langle 0| + \frac{2}{1+\Delta} Tr_B |t^-_it^+_{i+1}\rangle \langle 0| + \\
                                                      &&-\Delta Tr_B |0\rangle \langle t^0_it^0_{i+1}| + \frac{2}{1+\Delta}Tr_B |0\rangle \langle t^+_it^-_{i+1}| + \frac{2}{1+\Delta}Tr_B |0\rangle\langle t^-_it^+_{i+1}| ]\\
                                                      &=&- \frac{1}{2^L} \frac{J_{\parallel}}{4J_{\perp}} [2\Delta 4 S^z_{i} S^z_{i+1} + 4 \frac{2}{1+\Delta}( S^+_{i} S^-_{i+1}+h.c.)   ] \\
                                                      &=&- \frac{1}{2^L} \frac{4J_{\parallel}}{J_{\perp}(1+\Delta)} [\frac{1}{2} \Delta(1+\Delta)S^z_{i} S^z_{i+1} + \frac{1}{2}( S^+_{i} S^-_{i+1}+h.c.) ]
    \end{eqnarray*}
    Here we use the following relations:
    \begin{eqnarray*}
      Tr_B |t^0_it^0_{i+1}\rangle \langle 0| &=& \frac{1}{2^{L-2}}  [\langle \uparrow^B_i \uparrow^B_{i+1}  |t^0_i \rangle |t^0_{i+1} \rangle \langle s_i|\langle s_{i+1}| \uparrow^B_i \uparrow^B_{i+1}\rangle + \langle \downarrow^B_i \downarrow^B_{i+1}  |t^0_i \rangle |t^0_{i+1} \rangle \langle s_i|\langle s_{i+1}| \downarrow^B_i \downarrow^B_{i+1}\rangle + \\
                                             &&\langle \uparrow^B_i \downarrow^B_{i+1}  |t^0_i \rangle |t^0_{i+1} \rangle \langle s_i|\langle s_{i+1}| \uparrow^B_i \downarrow^B_{i+1}\rangle + \langle \downarrow^B_i \uparrow^B_{i+1}  |t^0_i \rangle |t^0_{i+1} \rangle \langle s_i|\langle s_{i+1}| \downarrow^B_i \uparrow^B_{i+1}\rangle]\\
                                             &=&  \frac{1}{2^L}[ |\downarrow^A_i \downarrow^A_{i+1}\rangle\langle \downarrow^A_i \downarrow^A_{i+1}| +
                                                 |\uparrow^A_i \uparrow^A_{i+1}\rangle\langle \uparrow^A_i \uparrow^A_{i+1}|
                                                 - |\uparrow^A_i \downarrow^A_{i+1}\rangle\langle \uparrow^A_i \downarrow^A_{i+1}|
                                                 - |\downarrow^A_i \uparrow^A_{i+1}\rangle\langle \downarrow^A_i \uparrow^A_{i+1}| ] \\
                                             &=& \frac{1}{2^L} 4 S^z_{i} S^z_{i+1}
    \end{eqnarray*}
    and
    \begin{eqnarray*}
      Tr_B |t^+_it^-_{i+1}\rangle \langle 0| &=& \frac{1}{2^{L-2}}  [\langle \uparrow^B_i \uparrow^B_{i+1}  |t^+_i \rangle |t^-_{i+1} \rangle \langle s_i|\langle s_{i+1}| \uparrow^B_i \uparrow^B_{i+1}\rangle + \langle \downarrow^B_i \downarrow^B_{i+1}  |t^+_i \rangle |t^-_{i+1} \rangle \langle s_i|\langle s_{i+1}| \downarrow^B_i \downarrow^B_{i+1}\rangle + \\
                                             &&\langle \uparrow^B_i \downarrow^B_{i+1}  |t^+_i \rangle |t^-_{i+1} \rangle \langle s_i|\langle s_{i+1}| \uparrow^B_i \downarrow^B_{i+1}\rangle + \langle \downarrow^B_i \uparrow^B_{i+1}  |t^+_i \rangle |t^-_{i+1} \rangle \langle s_i|\langle s_{i+1}| \downarrow^B_i \uparrow^B_{i+1}\rangle]\\
                                             &=& \frac{1}{2^L}[ -2|\uparrow^A_i \downarrow^A_{i+1}\rangle\langle \downarrow^A_i \uparrow^A_{i+1}| ] \\
                                             &=& \frac{1}{2^L} [ - 2 S^+_{i} S^-_{i+1} ]
    \end{eqnarray*}

    Third, we derive
    \begin{eqnarray*}
      Tr_B [|1\rangle\langle 1|] &=& \sum_{i,j}
                                     \big(\frac{J_{\parallel}}{4J_{\perp}}\big)^2 [\frac{2}{1+\Delta} | t^+_it^-_{i+1}\rangle  + \frac{2}{1+\Delta} | t^-_it^+_{i+1}\rangle - \Delta |t^0_i t^0_{i+1}\rangle]
                                     [\frac{2}{1+\Delta} \langle t^+_jt^-_{j+1}|  + \frac{2}{1+\Delta} \langle t^-_jt^+_{j+1}| - \Delta \langle t^0_j t^0_{j+1}|]\\
                                 &=&\big(\frac{J_{\parallel}}{4J_{\perp}}\big)^2 \sum_i
                                     [ \frac{2^2}{(1+\Delta)^2} Tr_B |t^+_it^-_{i+1}\rangle \langle t^+_{i+1}t^-_{i+2}| + \frac{2^2}{(1+\Delta)^2} Tr_B |t^+_it^-_{i+1}\rangle \langle t^-_{i+1}t^+_{i+2}| - \frac{2\Delta}{1+\Delta} Tr_B |t^+_it^-_{i+1}\rangle \langle t^0_{i+1} t^0_{i+2}| \\
                                 && \frac{2^2}{(1+\Delta)^2} Tr_B |t^-_it^+_{i+1}\rangle \langle t^+_{i+1}t^-_{i+2}| + \frac{2^2}{(1+\Delta)^2} Tr_B |t^-_it^+_{i+1}\rangle \langle t^-_{i+1}t^+_{i+2}| - \frac{2\Delta}{1+\Delta} Tr_B |t^-_it^+_{i+1}\rangle \langle t^0_{i+1} t^0_{i+2}| \\
                                 && - \frac{2\Delta}{1+\Delta} Tr_B |t^0_it^0_{i+1}\rangle \langle t^+_{i+1} t^-_{i+2}| - \frac{2\Delta}{1+\Delta} Tr_B |t^0_it^0_{i+1}\rangle \langle t^-_{i+1}t^+_{i+2}| +   \Delta^2 Tr_B |t^0_it^0_{i+1}\rangle \langle t^0_{i+1} t^0_{i+2}| ] \\
                                 && +\big(\frac{J_{\parallel}}{4J_{\perp}}\big)^2
                                    [ \frac{2^2}{(1+\Delta)^2} Tr_B |t^+_{i+1}t^-_{i+2}\rangle \langle t^+_{i}t^-_{i+1}| + \frac{2^2}{(1+\Delta)^2} Tr_B |t^+_{i+1}t^-_{i+2}\rangle \langle t^-_{i}t^+_{i+1}| - \frac{2\Delta}{1+\Delta} Tr_B |t^+_{i+1}t^-_{i+2}\rangle \langle t^0_{i} t^0_{i+1}| \\
                                 && \frac{2^2}{(1+\Delta)^2} Tr_B |t^-_{i+1}t^+_{i+2}\rangle \langle t^+_{i}t^-_{i+1}| + \frac{2^2}{(1+\Delta)^2} Tr_B |t^-_{i+1}t^+_{i+2}\rangle \langle t^-_{i}t^+_{i+1}| - \frac{2\Delta}{1+\Delta} Tr_B |t^-_{i+1}t^+_{i+2}\rangle \langle t^0_{i} t^0_{i+1}| \\
                                 && - \frac{2\Delta}{1+\Delta} Tr_B |t^0_{i+1}t^0_{i+2}\rangle \langle t^+_{i} t^-_{i+1}| - \frac{2\Delta}{1+\Delta} Tr_B |t^0_{i+1}t^0_{i+2}\rangle \langle t^-_{i}t^+_{i+1}| +   \Delta^2 Tr_B |t^0_{i+1}t^0_{i+2}\rangle \langle t^0_{i} t^0_{i+1}| ] \\
                                 &=& \frac{1}{2^{L-3}} \big(\frac{J_{\parallel}}{4J_{\perp}}\big)^2 [\frac{2^2}{(1+\Delta)^2}\frac{1}{2}(S^+_i S^-_{i+2}+h.c.)+\frac{\Delta^2}{8}4 2 S^z_i S^z_{i+2}] \\
                                 &=& \frac{1}{2^L} \big(\frac{J_{\parallel}}{J_{\perp}}\big)^2  \frac{1}{2}[\frac{2^2}{(1+\Delta)^2}\frac{1}{2}(S^+_i S^-_{i+2}+h.c.)+ \Delta^2 S^z_i S^z_{i+2}]
    \end{eqnarray*}
    And we need the relations:
    \begin{eqnarray}\label{}
      &&Tr_B | t^0_i t^0_{i+1} \rangle \langle t^0_{i+1} t^0_{i+2} | =  \frac{1}{8} 4 S^z_{i} S^z_{i+2} \\
      &&Tr_B | t^+_i t^-_{i+1} \rangle \langle t^-_{i+1} t^+_{i+2} | =  \frac{1}{2} | \uparrow^A_i \downarrow^A_{i+1} \downarrow^A_{i+2}\rangle \langle \downarrow^A_i \downarrow^A_{i+1}\uparrow^A_{i+2}| \\
      &&Tr_B | t^-_i t^+_{i+1} \rangle \langle t^+_{i+1} t^-_{i+2} | =  \frac{1}{2} | \downarrow^A_i \uparrow^A_{i+1}\uparrow^A_{i+2}\rangle \langle\uparrow^A_i \uparrow^A_{i+1} \downarrow^A_{i+2}| \\
      &&Tr_B | t^+_{i+1} t^-_{i+2} \rangle \langle t^-_{i} t^+_{i+1} | =  \frac{1}{2} | \uparrow^A_i \uparrow^A_{i+1} \downarrow^A_{i+2}\rangle \langle \downarrow^A_i \uparrow^A_{i+1}\uparrow^A_{i+2}| \\
      &&Tr_B | t^-_{i+1} t^+_{i+2} \rangle \langle t^+_{i} t^-_{i+1} | =  \frac{1}{2} | \downarrow^A_i \downarrow^A_{i+1}\uparrow^A_{i+2}\rangle \langle\uparrow^A_i \downarrow^A_{i+1} \downarrow^A_{i+2}| \\
    \end{eqnarray}

    At last, we sum up all calculations together:

    \begin{eqnarray}\label{}
      \rho_A &=& Tr_B[(|0\rangle+|1\rangle) (\langle 0|+\langle 1|)] \nonumber\\
             &=&\frac{1}{2^L} \big[ 1 - \frac{4J_{\parallel}}{J_{\perp}(1+\Delta)} [\frac{1}{2} \Delta(1+\Delta)S^z_{i} S^z_{i+1} + \frac{1}{2}( S^+_{i} S^-_{i+1}+h.c.) ]
                 +\big(\frac{J_{\parallel}}{J_{\perp}}\big)^2  \frac{1}{2}[\frac{2^2}{(1+\Delta)^2}\frac{1}{2}(S^+_i S^-_{i+2}+h.c.)+ \Delta^2 S^z_i S^z_{i+2}]   \big] \nonumber \\
             &\approx& \frac{1}{\mathcal{Z}}\exp(-\mathcal{H}^{per}_E)
    \end{eqnarray}
    , where
    \begin{align}\label{}
      \mathcal{H}^{\textit{per}}_E = \tilde J^{xy}_1 \sum_{i,\alpha=x,y} S^{\alpha}_i S^{\alpha}_{i+1} +  \tilde J^{zz}_1 \sum_{i} S^{z}_i S^{z}_{i+1}
      - \tilde J^{xy}_2 \sum_{i,\alpha=x,y} S^{\alpha}_i S^{\alpha}_{i+2} - \tilde J^{zz}_{2} \sum_i S^{z}_i S^{z}_{i+2},
    \end{align}
    and $\tilde J^{xy}_1=\frac{4}{1+\Delta}\frac{J_{\parallel}}{J_{\perp}}$,
    $\tilde J^{zz}_1=2 \Delta \frac{J_{\parallel}}{J_{\perp}}$,
    $\tilde J^{xy}_2=\frac{2}{(1+\Delta)^2}\big(\frac{J_{\parallel}}{J_{\perp}}\big)^2$ and
    $\tilde J^{zz}_2= \frac{\Delta^2}{2}\big(\frac{J_{\parallel}}{J_{\perp}}\big)^2$.
    Here we only keep the leading term in nearest neighbor and second nearest neighbor couplings.
    The form that we show in the main text is the case for isotropic case $\Delta=1$. The anisotropic
    form will be discussed in Sec.~\ref{app:aniso}.

\section{V. Comparison entanglement spectra with eigenvalues spectrum of parent Hamiltonian}
The equivalence between the entanglement Hamiltonian $\HE$ and parent Hamiltonian $H_P$
can be validated through the analysis of universal feature in the entanglement spectra (ES).
Here we show the comparison of ES of entanglement Hamiltonian
and eigenvalue spectrum of parent Hamiltonian in spin ladder model (the results for 1D spin chain model has been shown in the main text).
Fig.~\ref{fig:ES_K} shows typical ES (measured from the minimal value $\xi_0$) plotted
as a function of momentum $K=\frac{2\pi k}{L}$ ($k=0,1,...,L-1$),
since the translational symmetry along the chain direction is preserved.
For the isotropic case $\Delta=1$ (Fig.~\ref{fig:ES_K}(a)),
the low-lying excitations of ES form an arch structure,
which can be fitted by the des Cloiseaux-Pearson dispersion relations $\xi_i-\xi_0=v|\sin K|$ (red dashed line).
It strongly suggests the ES can resemble gapless quantum critical behavior which is intrinsic to
the quantum spin$-1/2$ Heisenberg chain \cite{Mikeska2004,Poilblanc2010}.
Importantly, the eigenvalue spectra of obtained $H_P$ shows the very similar features (Fig.~\ref{fig:ES_K}(b)).
As a direct comparison, we plot $\varepsilon_n-\varepsilon_0$ and $\xi_n-\xi_0$ ($\xi_n=-\log p_n$) in Fig.~\ref{fig:ES_K}(c).
It is found that eigenvalue $\varepsilon_n$ has one-to-one correspondence with $\xi_i$, and
a linear relationship $\varepsilon_n\propto \xi_n$ can be established (red dashed line).
Here, the comparison between entanglement spectra and eigenvalue spectra of $H_P$
clearly establishes the relationship between entanglement Hamiltonian and reduced density matrix: $\HE = f(H_P) \approx H_P$.

\begin{figure}[!htb]
\includegraphics[width=0.5\textwidth]{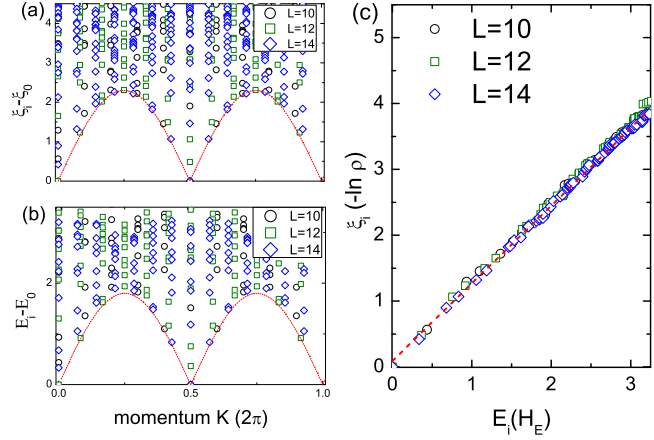}\\
\caption{(a) Entanglement spectra ($\xi_i-\xi_0$), obtained from reduced density matrix,
are grouped by total momentum $K$ along the chain direction.
(b) Energy spectra ($E_i-E_0$) of reconstructed entanglement Hamiltonian $\HE$.
In (a-b), the lowest spectra branch is fitted as $v|\sin K|$ by red dashed line.
(c) Direct comparison of entanglement spectra ($\xi_i-\xi_0$) and energy spectra ($E_i-E_0$).
All low-lying spectra are computed on $2\times L$ ladders
shown in black circles ($L=10$), green squares ($L=12$) and blue diamonds ($L=14$).
Here we set $\theta=\pi/3$ and $\Delta=1.0$.
} \label{fig:ES_K}
\end{figure}

\section{VI. Anisotropic Case}\label{app:aniso}
In the main text, we focus on the isotropic Heisenberg model. Here we briefly discuss the anisotropic case ($\Delta>1$).
In our extensive tests, our numerical scheme works well for both isotropic and anisotropic Heisenberg model.
For the anisotropic case, we can also map out the entanglement Hamiltonian within the same scheme.
Here we show spin ladder model (Fig. \ref{fig:model}(b)) and take $J_{\perp}/J_{\parallel}=4$ and $\Delta=2$ as an example:
\begin{eqnarray}
  &&\hat H= \hat H_A + \hat H_B +\hat H_{AB} \nonumber\\
  &&\hat H_{\alpha=A(B)}= J_{\parallel} \sum_{\langle ij\rangle} [S^x_{i,\alpha} S^x_{j,\alpha}+S^y_{i,\alpha} S^y_{j,\alpha} + \Delta S^z_{i,\alpha} S^z_{j,\alpha}] \nonumber\\
  &&\hat H_{AB}= J_{\perp} \sum_{i} [S^{x}_{i,A} S^{x}_{i,B} + S^{y}_{i,A} S^{y}_{i,B} + \Delta S^z_{i,A} S^z_{i,B}].
\end{eqnarray}
The targeting operator space is chosen to be:
\begin{eqnarray}
 &&\mathcal{H}_E= \sum^{N_r}_{n=1} J^{xy}_n \hat h^{xy}_n + J^{zz}_n\hat h^{zz}_n \nonumber\\
 &&\hat h^{xy}_n =\sum_{i=1}^{L} (S^x_{i} S^x_{i+n} +S^y_{i} S^y_{i+n}) ,\,\,\,
  \hat h^{zz}_n = \sum_{i=1}^{L}  S^z_{i} S^z_{i+n}.
\end{eqnarray}
Within the same scheme shown in the main text, the obtained parameters of entanglement Hamiltonian is shown in Tab.~\ref{tab:XXZ}.
As shown in Tab.~\ref{tab:XXZ}, $\HE$ breaks the spin rotation symmetry $J^{xy}_n\neq J^{zz}_n$.
We also confirm that the parameters in $\HE$ can be compared with perturbation theory
as shown in the Sec. \ref{app:perturbation}.
These facts point to that  $\HE$ is effectively described by
the XXZ spin chain with spin rotation symmetry breaking.
This is not surprising since the parent Hamiltonian breaks spin rotation symmetry explicitly.

\begin{table}[!htb]
  \caption{Parameters of entanglement Hamiltonian $\HE$ for anisotropic spin ladder model.
    Here we set $J_{\perp}/J_{\parallel}=4$ and $\Delta=2$. } \label{tab:XXZ}
  \begin{tabular}{c|c|c|c|c|c}
    \hline
    \hline
    $L$&$g_0$&$J^{xy}_1$&$J^{xy}_2$&$J^{zz}_1$&$J^{zz}_2$\\
    \hline
    $10$ & $1.34\times 10^{-8}$& $0.304$ & $-0.039$ & $0.952$ & $-0.065$ \\
    $12$ & $6.37\times 10^{-7}$& $0.299$ & $-0.039$ & $0.951$ & $-0.061$ \\
    $14$ & $8.89\times 10^{-7}$& $0.303$ & $-0.038$ & $0.950$ & $-0.067$ \\
    \hline
  \end{tabular}
\end{table}

  \end{appendices}

\end{widetext}

\end{document}